# Ultrafast laser inscription: an enabling technology for astrophotonics

**Robert R. Thomson, Ajoy K. Kar**

*School of Engineering and Physical Sciences, Heriot-Watt University, Edinburgh, EH14 4AS, UK.*
*http://nlo.eps.hw.ac.uk/*

**Jeremy Allington-Smith***

*Centre for Advanced Instrumentation, Physics Dept ,Durham University, Durham DH1 3LE, UK*
*\*Corresponding author J.R.Allington-Smith@durham.ac.uk*

**Abstract:** The application of photonics to astronomy offers major advantages in the area of highly-multiplexed spectroscopy, especially when applied to extremely large telescopes. These include the suppression of the near-infrared night-sky spectrum [J. Bland-Hawthorn et al, Opt. Express **12**, 5902 (2004), S. G. Leon-Saval et al, Opt. Lett. **30**, 2545 (2005)] and the miniaturisation of spectrographs so that they may integrated into the light-path of individual spatial samples [J. Bland-Hawthorn et al, Proc SPIE 6269, 62690N (2006)]. Efficient collection of light from the telescope requires multimode optical fibres and three-dimensional photonic devices. We propose ultrafast laser inscription (ULI) [R. R. Thomson et al, Opt. Express **15**, 11691 (2007)] as the best technology to fabricate 3D photonic devices for astrophotonic applications.

## References and links

## 1. Highly multiplexed Spectroscopy

Highly Multiplexed Spectroscopy (HMS) is a key observational technique for the next generation of Extremely Large Telescopes (ELTs). HMS includes both simultaneous observations of multiple faint objects at the limits of detection (multiple object spectroscopy: MOS) and spatially-resolved spectroscopy over contiguous fields of brighter structured objects (integral field spectroscopy: IFS) and a mixture of the two (diverse field spectroscopy: DFS). Examples of MOS include studies of large samples of distant galaxies to determine the cosmic equation of state; and individual stars in this and nearby galaxies to determine their origins by "galactic archaeology". Examples of IFS include studies of the nuclei of active galaxies which harbour supermassive black holes and uncovering evidence of galactic mergers. DFS is aimed mainly at the assembly of primeval galaxies and studies of dense star clusters where intermediate mass black holes may reside. These all require multiplex factors of $10^2 \rightarrow 10^5$. With ELTs the sensitivity is boosted by both the increase in collecting area and by the improved spatial resolution available, reinforcing the importance of HMS for the future.

The largest multiplex gains for MOS are generally achieved with fibre systems in which optical fibres are placed at the focal surface of the telescope to intercept light from the targets of interest. These use conventional step-index multimode fibres (MMFs) made from fused silica with a core diameter of ~100μm. They are convenient for large scale integration due to their flexibility but are compromised in efficiency by étendue non-conservation and high attenuation at wavelengths <400nm. Slit masks can be used with wide-field spectrographs to improve efficiency but this also restrict the choice of targets. IFS can also be realized using composite fibre-lenslet and monolithic diamond-machined technologies [1].

## 2. Photonics applied to highly-multiplexed spectroscopy

*2.1 Background suppression*

Removal of the night-sky spectrum is a critical requirement for NIR spectroscopy. The spectrum consists of an aperiodic forest of very narrow, intense emission lines which vary with both position in the sky and time. This degrades the signal/noise ratio (SNR) of the object spectrum or requires costly or lossy processes to remove them. Removal is best done before the light enters the spectrograph because the optics and disperser add significant wings to the spectral PSF [2] which hinders the removal of the background.

A promising photonic solution is couple the light from the telescope into an optical fibre inscribed with an Aperiodic Fiber Bragg-Grating (AFBG). AFBGs on single-mode fibres have already been demonstrated to suppress the brightest OH lines at 1500-1570nm [3] but, for astronomy, the celestial light must be captured using MMFs to increase the coupling efficiency. Since each mode has a different propagation constant, the Bragg condition in a MMF-FBG is satisfied at different wavelengths for each transverse mode. This leads to an error in the passband which degrades the spectrum. To overcome this problem, a novel photonic device could be used to couple the multimode light to an array of SMFs with inscribed AFBGs [4] as discussed further in §3.

*2.2 Photonic spectrographs*

The trend towards increased multiplex gain in astronomy and the increased physical size of the telescopes, allied to the unfavorable scaling laws for the instruments, make fibre-based solutions attractive for ELTs [1]. However, the scaling laws imply impractically large instruments unless the effective slit width can be reduced so that the required resolving power can be obtained with a smaller disperser. This requires high image quality over the wide fields necessary for HMS which may not be delivered by AO initially Even then, observations of faint galaxies need wider slits to match their characteristic physical size.

The main problem is that traditional dispersers (surface-relief or index-modulated) have an inherent geometric constraint that makes size reduction difficult. To generalize previous discussions of photonic spectrographs [5], consider the spectral resolution of a spectrograph with a slit width sized to optimize SNR and therefore generally wider than the diffraction limit of the telescope. The spectral resolution is then determined by the smallest resolvable angle emerging from the disperser, $\delta\beta$, via Eq. 1:

$$\delta\lambda = \left[\frac{d\lambda}{d\beta}\right]\delta\beta = \frac{s}{\Gamma f_S}\left[\frac{d\lambda}{d\beta}\right] = \chi \frac{D_T}{D_S}\frac{1}{\Gamma}\left[\frac{d\lambda}{d\beta}\right] \qquad \text{Eq. (1)}$$

using conservation of étendue $\delta\beta = s'/f_C = s/f_S$ where $s$ and $s'$ are the widths of the slit and the image of it on the detector respectively and $f_S$ and $f_C$ are the focal lengths of the collimator and camera respectively. $\Gamma$ is the anamorphic factor introduced the disperser and the term in square brackets is the angular dispersion. In terms of the angular width of the slit projected on the sky, $\chi = sD_S/f_SD_T$, where $D_T$ is the telescope aperture diameter and $D_S$ is the diameter of the collimated beam incident on the disperser, the resolving power is given by Eq. (2):

$$R \equiv \frac{\lambda}{\delta\lambda} = \Gamma \frac{D_S}{D_T}\frac{\lambda}{\chi}\left[\frac{d\lambda}{d\beta}\right]^{-1} \qquad \text{Eq. (2)}$$

This is a very general expression, independent of the details of the disperser, and dependent only on the slitwidth, size of the collimated beam and the anamorphic factor (~1).

For a spectrograph using a diffraction grating, the dispersion is obtained from Eq. (3) which gives the interference between beams emerging from adjacent scattering centers in a medium with refractive index, $n$, at angle $\beta$:

$$n(\sin\alpha + \sin\beta) = m\rho\lambda \qquad \text{Eq. (3)}$$

where the input angle is $\alpha$ and $\rho$ is the linear density of scattering centers in the dispersion direction (i.e. the rulings). The dispersion and anamorphic factor respectively are $\cos\beta/(m\rho)$ and $\cos\beta/\cos\alpha$ (for air), so from Eq. (2) the resolving power is given by Eq. (4)

$$R_G = \frac{D_S}{\chi D_T}\left(\frac{\sin\alpha + \sin\beta}{\cos\alpha}\right) = \frac{2D_S \tan\gamma}{\chi D_T} \qquad \text{Eq. (4)}$$

where we have substituted from Eq. (3) to eliminate details of the disperser in the first expression. The second expression is for the blaze condition when $\alpha = \beta \equiv \gamma$. This occurs when the scattering centers are in phase with the edges of their clear apertures to maximize the intensity from each facet. The geometry of the optical system imposes a practical restriction that $0 < \tan\gamma < 4$.

So far, this is very familiar to astronomers. However, if we replace the grating by an array of waveguides – defined generically to include SMFs – with path differences between adjacent scattering centers of $\Delta y = (dy/dx)\Delta x$ where $x$ is the coordinate along the fibre array and $\Delta x$ is the centre-centre pitch, then Eq. (3) becomes Eq. (5)

$$[dy/dx] + \sin\beta = m\rho\lambda \qquad \text{Eq. (5)}$$

where $\rho = 1/\Delta x$. Thus the dispersion and the anamorphic factor respectively are $\cos\beta/(m\rho)$ and unity. So, the resolving power, from Eq. (2), is given by Eq. (6)

$$R_P = \frac{D_S}{\chi D_T}\left(\left[\frac{dy}{dx}\right] + \sin\beta\right)\frac{1}{\cos\beta} = \frac{D_S}{\chi D_T}\left[\frac{dy}{dx}\right] \qquad \text{Eq. (6)}\backslash$$

since the blaze condition occurs when the phase difference between the centre and edge of a fibre, $(\pi/\lambda)\sin\beta$, vanishes, i.e. when $\beta = 0$. This is determined only by the exit angle since the phase is constant over each waveguide exit aperture

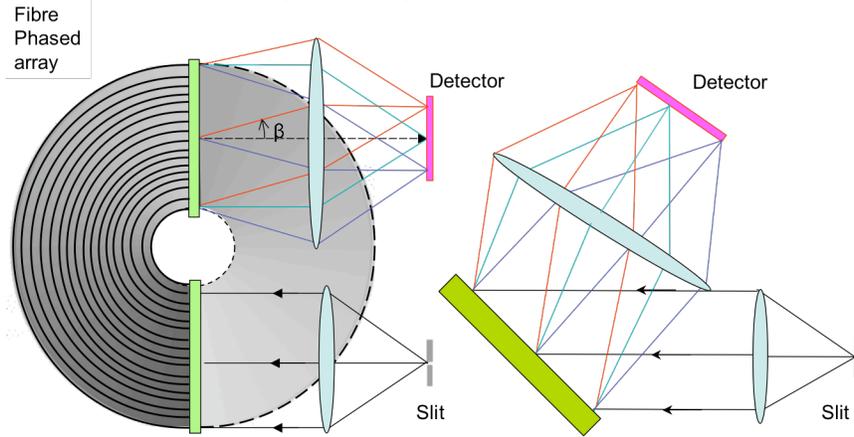

Fig. 1: Left - Cartoon of photonic spectrograph with a phased waveguide array as the disperser. Right – traditional spectrograph of the same overall dimensions, but worse spectral resolution.

Compared with the grating spectrograph, there is no limit imposed by geometry, only by the construction of the waveguide array. This means that the resolving power can exceed that of a grating spectrograph by a factor $W = [dy/dx]/2\tan\gamma$ which can be made arbitrarily large as demonstrated in Fig. 1 where the waveguides are wound $N$ times round a circle. For $N = 6$ windings, the advantage is $W = (2N+1)\pi/8 = 5$ in a volume comparable to a traditional spectrograph. Conversely this means that, for a given resolving power, the characteristic size of the spectrograph may be drastically reduced (by a factor $\sim W^2$) since the disperser size dominates the spectrograph dimensions.

### 3. Coupling celestial light into photonic devices

Light arriving at a telescope from the depths of the cosmos is inherently multimoded. Furthermore, the sample size is determined by the SNR of the required data product (e.g. line flux, radial velocity) rather than the achievable spatial resolution, and this is often larger than the mode area or core size of a SMF. Coupling light directly into a SMF is inefficient [6] so it is better to use MMFs where losses can be limited to $\sim$1%. However, existing photonic devices (e.g. FBGs and Arrayed Waveguide Gratings: AWGs), have been developed for single-mode input. This presents a major challenge to astrophotonics.

Once the light from the telescope has been collected by a MMF, it must be processed by the photonic device. Since each transverse mode in a MMF exhibits a different propagation constant, a different electric field distribution and a different phase profile across the mode, careful control and manipulation of these modes is essential to optimise the operation of the photonic device. For the OH-suppression (OHS) application (§2), each transverse mode supported by a MMF has a slightly different propagation constant, which degrades the rejection passband and reduces performance. To overcome this, it may be possible to use optical fibre tapering techniques to create a low-loss transition between a MMF and an array of SMFs [4] such that the transverse modes supported by the MMF may be transformed into the supermodes supported by the SMF array. Furthermore, since the supermodes supported by the SMF array are degenerate, with each exhibiting an almost identical propagation constant, a multimode fibre filter with the spectral characteristics of a SMF-FBG could be realized by placing Bragg gratings in the SMFs. Such a device would remove the night-sky background with high coupling efficiency and throughput.

The throughput of the proof-of-concept device demonstrated in ref. [4] was limited to $\sim$3%. The authors state that this was due to the mismatch in the number of modes supported by the MMF and the number of supermodes supported by the array of SMFs. Although this may be reduced by increasing the number of SMFs in order to address the full modal content of the incoming MMF, scaling this technology to MMFs that support 50-100 modes will require greater complexity in design and manufacture. Furthermore for HMS, each of the $10^2$-$10^5$ MMFs would need to be adapted in this way implying up to $\sim 10^7$ individual SMFs – an integrated solution is clearly required.

### 4. Technology options

*4.1 Requirements*

From the previous discussion we can define two critical astrophotonic applications of particular interest to HMS: firstly, the suppression of atmospheric OH emission lines using AFBGs and, secondly, the miniaturization and modularization of spectrographs so that, in the limit, each MMF feeds an individual spectrograph. Both of these applications will require the manipulation of highly multimode light and thus integrated and advanced 3D mode converters. In addition, the photonic spectrograph requires a disperser fashioned as a phased array of waveguides with arbitrarily large phase shifts between adjacent scattering centers. Furthermore, we anticipate that the implied multiplexity of the waveguides (50-100 for the OHS application alone) will require the ability to weave the waveguides in 3D to allow low-loss and low cross-talk waveguide crossovers – topologically impossble in planar devices..

*4.2 Ultrafast Laser Inscription (ULI)*

Although optical fibre technology could be used to achieve these goals on a small scale, it may not be suitable for the large-scale production of many thousands of individual modules coupled to individual MMFs. We suggest that a more suitable technology may be the revolutionary technique of ultrafast laser inscription (ULI).

Ultrashort laser pulses are the shortest events ever created by humanity. Due to their brevity, extremely high peak powers can be obtained using relatively low energy pulses and average powers, e.g. a 100 fs pulse of 1.0 µJ energy peaks at a power of 10MW. When focused inside a dielectric material that is normally transparent to the laser wavelength, the peak intensity can be high enough to induce various nonlinear mechanisms that deposit optical energy in the material at the focus [7]. The deposited energy can induce highly localised modifications to the structure of the material which can in turn manifest themselves in a variety of ways, such as changes in the refractive index [8] or susceptibility to chemical etching [9] of the modified material. Since the light-matter interaction is nonlinear, only occurring above an intensity threshold, 3D structures can be directly inscribed inside the material by translating the material in 3D through the focus of the laser beam.

3D photonic structures can be directly inscribed inside the material. For example, the refractive index change can be used to directly inscribe 3D optical waveguides [10, 11], and optical waveguides have already been demonstrated that simultaneously exhibit low single mode fibre-to-waveguide coupling losses (~0.1 dB/facet) and propagation losses (0.3 dB/cm) [12]. The etch-rate modification on the other hand can be used to create micro-optics [13], free-standing multimode optical waveguides [9] and high aspect ratio micro-fluidics [14]. Thus, ULI is a revolutionary new 3D photonic device fabrication technology that provides the potential to realise devices that would have otherwise been impossible using conventional photonic device fabrication technologies.

*4.3 Astrophotonic applications of ultrafast laser inscription*

ULI can be used to make photonic devices to address the astrophotonic applications highlighted above. These would be very challenging to create initially, but once mastered the technique could be scaled to fabricate thousands of devices.

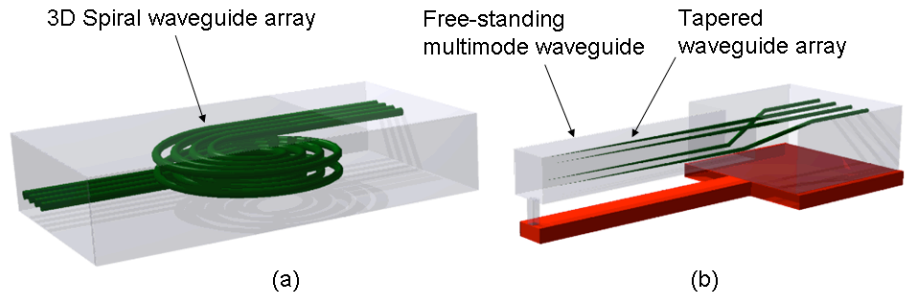

Fig. 2: Conceptual diagrams of (a) a highly dispersive 3D waveguide array and (b) an integrated OH-fluorescence filter. In both (a) and (b), the inscribed waveguides are green.

Figure 2 indicates how a ULI fabricated highly dispersive waveguide array and OHS device might appear. The highly dispersive waveguide array could be fabricated by directly inscribing a 3D spiral array of waveguides in a glass substrate using the refractive index change induced by the ULI process. The OHS device on the other hand could be fabricated by irradiating the substrate in such a way that after etching, a free-standing multimode optical waveguide would be formed, similar to that described in [9]. Before the etching was performed however, a tapered array of optical waveguides could be inscribed inside the multimode waveguide. Light from the telescope would be coupled into the multimode waveguide, the transverse modes of the multimode waveguide would then adiabatically transform into the supermodes supported by the waveguide array. The OH-fluorescence could then be filtered from the celestial light using aperiodic gratings inscribed in the waveguides by modulating the power during the inscription. After filtering, the waveguides could flare out into a linear array, for coupling the filtered light into a conventional or photonic spectrograph.

The power modulation technique has already been used to fabricate monolithic waveguide lasers [15, 16] incorporating distributed feedback gratings with refractive index modulations in excess of $10^{-4}$ [16]. This level of index modulation is certainly adequate for the fabrication of complex grating structures [3].

*4.4 Technological challenges for ultrafast laser inscription*

For ULI to realise its full potential for astrophotonic applications, a number of technological issues must first be addressed. The first of these is the relatively high propagation losses currently associated with ULI fabricated waveguides. The lowest propagation loss reported to date for a ULI fabricated waveguide is ~0.1 dB/cm at 1.55μm wavelength[17]. This is an order of magnitude higher than can be achieved using more conventional waveguide fabrication technologies and, depending on the wavelength of the light being used, limits the maximum waveguide length to a few tens of cm. It should be noted however that although the ULI field has grown rapidly over the last decade, ULI is still a relatively immature technology. It is only now that researchers are beginning to study and understand the origins of the propagation losses in ULI fabricated waveguides [18], investigate exactly how the various ULI parameters affect the waveguide propagation losses [19], and find ways to reduce them. Consequently, we see no reason why, in the future, the propagation losses of ULI fabricated waveguides cannot be reduced to levels that will be acceptable for astrophotonic devices that require long waveguide lengths.

The second, and probably the more challenging, technological issue that must be addressed is the low refractive index contrast of ULI fabricated waveguides. Although it appears ULI can be used to inscribe waveguides in almost any transparent dielectric material, most commonly used substrate materials, such as silicate, borosilicate and phosphate glasses, exhibit modest (< 0.5 %) refractive index changes under irradiation. This limits the bend radii of the inscribed waveguides to ~50mm before radiation losses become unacceptable [12, 20]. Clearly, this is a limitation that must be addressed as it increases the device footprint and limits the complexity of the devices that can be created.

Currently, a number of routes show promise for overcoming this limitation. For example, Sugioka's group at Riken have already demonstrated low-loss (< 0.3 dB) 90° waveguide bends using total-internal-reflection micro-mirrors [21]. These mirrors were created using ULI by irradiating regions that were then selectively removed using the etching technique described in section 4.2. Another route may be to use more exotic substrate materials that exhibit larger refractive index changes under irradiation. The Knox group at the University of Rochester have, for example, recently fabricated waveguides in a hydrogel polymer with a refractive index contrast of $\approx$ 4 % [22], and demonstrated waveguides with 1.0 mm bend radii that exhibited radiation losses below 0.1 dB/cm for 632 nm light. The Hirao group at Kyoto University have also recently demonstrated that silicon-rich structures can be directly inscribed in silicate glass using femtosecond laser pulses [23]. Since the refractive index of silicon is $\approx$ 3.4, techniques such as this could pave the way to extremely high refractive index contrast waveguides operating in the IR with very small bend radii.

It is our belief that techniques such as these will be optimised to a level that will enable the inscription of low-loss waveguides with tight bends, thus enabling the fabrication of highly integrated 3D astrophotonic devices.

## 5. Conclusions

In this paper we have proposed for the first time that ULI will be an enabling technology for astrophotonics. We currently see no other technology that can provide the unique fabrication capabilities of ULI.